\documentclass[aps,showpacs,preprintnumbers,amsmath,amssymb]{revtex4}

\UseRawInputEncoding
\usepackage{dcolumn}
\usepackage[dvips]{epsfig}

\usepackage{float}
\usepackage{graphicx}
\usepackage{epstopdf}
\usepackage{graphicx}
\usepackage{epstopdf}
\usepackage{color}
\usepackage{subfig}
\usepackage{amsmath,amssymb,amsfonts}

\begin{document}
\baselineskip=0.8 cm
\title{\bf Quasinormal modes of  a scalar perturbation around a rotating BTZ-like black hole in Einstein-bumblebee gravity}

\author{Chengjia Chen$^{1,2}$, Qiyuan Pan$^{1,2,3}$\footnote{panqiyuan@hunnu.edu.cn}, and Jiliang Jing$^{1,2}$\footnote{jljing@hunnu.edu.cn}}

\affiliation{$^{1}$Department of Physics, Key Laboratory of Low Dimensional Quantum Structures and Quantum Control of Ministry of Education, and Synergetic Innovation Center for Quantum Effects and Applications, Hunan Normal University, Changsha, Hunan
410081, China}
\affiliation{$^{2}$Institute of Interdisciplinary Studies, Hunan Normal University, Changsha, Hunan 410081, China}
\affiliation{$^{3}$ Center for Gravitation and Cosmology, College of Physical Science and Technology, Yangzhou University, Yangzhou 225009, China}

\begin{abstract}
\baselineskip=0.6 cm
\begin{center}
{\bf Abstract}
\end{center}

We analytically study the quasinormal modes of a scalar perturbation around a rotating BTZ-like black hole in the Einstein-bumblebee gravity. We observe that the Lorentz symmetry breaking parameter imprints only in the imaginary parts of the quasinormal frequencies for the right-moving and left-moving modes. The perturbational field decays more rapidly for the negative Lorentz symmetry breaking parameter, but more slowly for the positive one. The forms of the real parts are the same as those in the usual BTZ black holes. Moreover, we also discuss the $AdS/CFT$ correspondence from the quasinormal modes and  find that  the Lorentz symmetry breaking parameter enhances the left and right conformal weights $h_L$ and $h_R$ of the operators dual to the scalar field in the boundary. These results could be helpful to understand the $AdS/CFT$ correspondence and the Einstein-bumblebee gravity with the Lorentz symmetry violation.

\end{abstract}

\maketitle
\newpage

\section{Introduction}

Lorentz invariance, as a fundamental symmetry, has been of great importance in both Einstein's theory of general relativity and the standard model of particle physics. However, the recent development of unified gauge theories and the observed signals from high energy cosmic rays \cite{lv01,lv02} imply that the Lorentz symmetry may spontaneously break at a higher scale of energy. Thus, studying the Lorentz violation may be a useful way to obtain a deeper understanding of nature. There are many theories contained the Lorentz violation, for example, the standard model extension \cite{lv02gc,lv02gc1,lv02gc2}, string theory \cite{string1}, and so on.

One of simple effective theories with the Lorentz violation is the so-called Einstein-bumblebee gravity \cite{lvbh2}. In this model,  the spontaneous breaking of Lorentz symmetry is induced by a nonzero vacuum expectation value of a bumblebee vector field $B_{\mu}$ with a suitable potential. The corresponding effects of the Lorentz violation on  black hole physics and  cosmology  have been extensively studied in  \cite{lvbh2s1,lvbhh1,lvbhh2,lvbhh12,lvbhh3,lvbhh4,lvbhh5,lvbhh6,lvbhh7,lvbhh8,lvbh10,ReyesSS,UniyalKS,KhodadiS,Fang}. The first black hole solution in such a theory of Einstein-bumblebee gravity is obtained by Casana \emph{et al.}  \cite{lvbh1}, which is an exact solution describes the gravity of a static neutral black hole. The potential observing effects of the Lorentz violation imprinted in the classical tests including the gravitational lensing \cite{lvbh3} and quasinormal modes \cite{lvbh5} have been investigated in this black hole spacetime. The influence of the Lorentz violation on the Hawking radiation \cite{lvbh4} is also studied for the black hole. Moreover, other spherically symmetric black hole solutions (containing global monopole \cite{lvbh6}, cosmological constant \cite{lvbh7}, or Einstein-Gauss-Bonnet term \cite{lvbh8}) and  the traversable wormhole solution \cite{lvbh9} have been found in the framework of the bumblebee gravity theory. Furthermore, the rotating black hole solution in the Einstein bumblebee gravity \cite{lvbhrot1} is also obtained, and the information about the Lorentz violation stored in black hole shadow \cite{lvbhrot1,lvbhrot1s}, accretion disk \cite{lvbhrot2}, superradiance of black hole \cite{lvbhrot3} and particle's motion \cite{lvbhrot4} is discussed.  A Kerr-Sen-like black hole with a bumblebee field has also been investigated \cite{lvbhrot5}.  With the observation data of quasi-periodic oscillations frequencies of GRO J1655-40, XTE J1550-564, and GRS 1915+105 \cite{RPM1, XTE1,GRS1}, the range of the Lorentz symmetry breaking parameter is also constrained for the rotating black hole in the Einstein-bumblebee theory of gravity \cite{test1}. These investigations are useful for detecting the effects arising from the Lorentz symmetry breaking induced by the bumblebee vector field.

Three-dimensional gravity has been investigated extensively because it certainly offers potential
insights into quantum gravity. Recently, a three-dimensional rotating BTZ-like black hole solution was obtained in the Einstein-bumblebee gravity \cite{d2302}. It was found that the bumblebee field doesn't affect the locations of the black hole horizon and ergosphere. In this paper, we will study the quasinormal modes of a scalar perturbation around such a rotating BTZ-like black hole. The main motivation is to probe the effects of the Lorentz symmetry breaking arising from the bumblebee vector field on the quasinormal modes and the stability of black hole under the scalar perturbation. Moreover, the $AdS_3/CFT_2$ correspondence \cite{AdSCFT1,AdSCFT2,AdSCFT3} indicates that there exists a dual between the asymptotical $AdS_3$ gravity in the bulk and the two-dimensional conformal field theory in the boundary. Although the $AdS/CFT$ correspondence of this black hole has been discussed and the central charges of the dual CFT on the boundary have been computed \cite{d2302}, it is still an open issue how the Lorentz symmetry breaking affects the left and right conformal weights $h_L$ and $h_R$ of the operators dual to the perturbational field in the boundary. Thus,  we will further study  this interesting issue on the $AdS_3/CFT_2$ correspondence in the Einstein-bumblebee gravity from quasinormal modes.

The plan of our paper is organized as follows. In Sec. II, we will review briefly the rotating BTZ-like black hole in the Einstein-bumblebee gravity. In Sec. III, we will study the quasinormal modes of a scalar perturbation around the rotating BTZ-like black hole. Our results indicate that the presence of such Lorentz symmetry breaking modifies the behavior of the quasinormal modes in the black hole and changes the left and right conformal weights $h_L$ and $h_R$ of the operators dual to the scalar field in the boundary. Finally, in the last section we will include our conclusions.

\section{Rotating BTZ-like black hole in Einstein-bumblebee gravity}

In this section, we review briefly the rotating BTZ-like black hole in the Einstein-bumblebee gravity obtained in \cite{d2302}. The action of Einstein-bumblebee gravity with a negative cosmological constant $\Lambda=-\frac{1}{l^2}$ in the three dimensional spacetime can be expressed as \cite{string1,lvbh2,lvbh2s1,lvbhh1,lvbhh2,lvbhh12,lvbhh3,lvbhh4,lvbhh5,lvbhh6,lvbhh7,lvbhh8,lvbh10}
\begin{eqnarray}\label{action}
S=\int d^3x\sqrt{-g}\bigg[\frac{1}{2\kappa}\bigg(R+\frac{2}{l^2}\bigg)+\frac{\xi }{2\kappa} B^{\mu\nu}R_{\mu\nu}-\frac{1}{4}B^{\mu\nu}B_{\mu\nu}-V(B_{\mu}B^{\mu}\mp b^2)\bigg],
\end{eqnarray}
where $R$ is the Ricci scalar and $\kappa$ is a constant related to the three-dimensional
Newton's constant $G$ by $\kappa=8\pi G$. The coupling constant $\xi$ has the dimension of $M^{-1}$ and the strength of the bumblebee field $B_{\mu}$ is $B_{\mu\nu}=\partial_{\mu}B_{\nu}-\partial_{\nu}B_{\mu}$. In this theoretical model of the bumblebee field, the potential $V$ has a minimum at $B_{\mu}B^{\mu}\pm b^2=0$ (where $b$ is a real positive constant), which yields a nonzero vacuum value $\langle B_{\mu}\rangle=b_{\mu}$ with $b_{\mu}b^{\mu}=\mp b^2$ and leads to the destroying of the $U(1)$ symmetry. The  signs ``$\pm$" in the potential  determine whether the field $b_{\mu}$ is timelike or spacelike. The nonzero vector background $b_{\mu}$ spontaneously gives rise to the Lorentz symmetry violation \cite{lvbh1,lvbh2, lvbh2s1,lvbhh1,lvbhh2,lvbhh12,lvbhh3,lvbhh4,lvbhh5,lvbhh6,lvbhh7,lvbhh8,lvbh10}.

Varying the action (\ref{action}) with respect to the metric, one can find that the extended vacuum Einstein equation in the Einstein-bumblebee gravity (\ref{action}) becomes \cite{string1,lvbh2, lvbh2s1,lvbhh1,lvbhh2,lvbhh12,lvbhh3,lvbhh4,lvbhh5,lvbhh6,lvbhh7,lvbhh8,lvbh10}
\begin{eqnarray}\label{dmmass}
R_{\mu\nu}-\frac{1}{2}g_{\mu\nu}R=T_{\mu\nu},
\end{eqnarray}
with
\begin{eqnarray}
T_{\mu\nu}&=&B_{\mu\alpha}B_{\;\nu}^{\alpha}-g_{\mu\nu}\bigg(\frac{1}{4}B_{\alpha\beta}B^{\alpha\beta}+V\bigg)+2B_{\mu}B_{\nu}V'
+\frac{\xi}{\kappa}\bigg[\frac{1}{2}g_{\mu\nu}B_{\alpha}B^{\alpha}-B_{\mu}B^{\alpha}R_{\alpha\nu}-B_{\nu}B^{\alpha}R_{\alpha\mu}
\nonumber\\
&+&\frac{1}{2}\nabla_{\alpha}\nabla_{\mu}(B^{\alpha} B_{\nu})+\frac{1}{2}\nabla_{\alpha}\nabla_{\nu}(B^{\alpha} B_{\mu})-\frac{1}{2}\nabla^2(B_{\mu} B_{\nu})-\frac{1}{2}g_{\mu\nu}\nabla_{\alpha}\nabla_{\beta}(B^{\alpha} B^{\beta})\bigg].
\end{eqnarray}
The extended Einstein equation (\ref{dmmass}) admits a rotating BTZ-like black hole solution with a metric \cite{d2302}
\begin{eqnarray}\label{metric}
ds^{2}&=&-f(r')dt^{2}+\frac{(s+1)}{f(r')} dr'^{2}+r'^2 \left(d\theta-\frac{j}{2r'^2}dt\right)^2,
\end{eqnarray}
where
\begin{eqnarray}
f(r')=\frac{r'^2}{l^2}-M+\frac{j^2}{4r'^2}.
\end{eqnarray}
Here $M$ and $j$ are the mass and the spin parameters of the black hole (\ref{metric}), respectively. The bumblebee field has the form $b_{\mu}=(0,b\xi, 0)$  and the parameter $s=\xi b^2$ describes the spontaneous Lorentz symmetry breaking arising from the Einstein-bumblebee vector field.  The determinant of the metric (\ref{metric}) is $g=-(s+1)r'^2 $, which means that the metric becomes degenerate when $s=-1$. In addition, the Kretschmann scalar is \cite{d2302}
\begin{eqnarray}
R^{\mu\nu\rho\sigma}R_{\mu\nu\rho\sigma}=\frac{12}{l^4(1+s)^2}.
\end{eqnarray}
It is clear that the spacetime (\ref{metric}) is singular as $s=-1$. Thus, in order to maintain that there is no curvature singularity in the spacetime (\ref{metric}) and to conveniently compare with the case without the Lorentz symmetry breaking (i.e., $s=0$), we here restrict  the parameter $s$  in the range of $s>-1$, and then the coupling $\xi$ is limited in the region $\xi>-\frac{1}{b^2}$. The horizons of the black hole (\ref{metric}) are located at
\begin{eqnarray} \label{shjr}
r'^2_{\pm}=\frac{1}{2}\left(Ml^2\pm l \sqrt{M^2l^2-j^2}\right),
\end{eqnarray}
where the signs ${+}$ and $-$ correspond to the outer and inner horizons, respectively. It is easy to obtain that the horizon radius of the black hole does not depend on the spontaneous Lorentz symmetry breaking parameter $s$. However, the Hawking temperature of the event horizon \cite{d2302}
\begin{eqnarray}
T_H=\frac{1}{2\pi\sqrt{s+1}}\left(\frac{r'_+}{l^2}-\frac{j^2}{4r'^3_+}\right),
\label{hawkingt}
\end{eqnarray}
is related to the breaking parameter $s$ and decreases with this parameter. Substituting the relationship $\frac{1}{l^2}=-\frac{\Lambda}{3}$ and $j=8J$ (where $J$ is the angular momentum of the black hole), and replacing the parameter $s$ by $\ell$, the Hawking temperature (\ref{hawkingt}) reduces to its form obtained in Ref. \cite{d2302}.

\section{Quasinormal modes of a scalar perturbation around a rotating BTZ-like black hole in Einstein-bumblebee gravity}

To analytically obtain the quasinormal modes of a scalar perturbation, we introduce a radial coordinate transformation
 \begin{eqnarray}\label{transfor}
r'\rightarrow r=r'^2,
\end{eqnarray}
 and the form of the metric (\ref{metric}) becomes
\begin{eqnarray}\label{metric2}
ds^{2}&=&-f(r)dt^{2}+\frac{(s+1)}{4rf(r)} dr^{2}+r \left(d\theta-\frac{j}{2r}dt\right)^2,
\end{eqnarray}
where
\begin{eqnarray}
f(r)=\frac{r}{l^2}-M+\frac{j^2}{4r}.
\end{eqnarray}
The horizon radii of the black hole (\ref{metric2}) become
\begin{eqnarray} \label{shjr}
r_{\pm}=\frac{1}{2}\left(Ml^2\pm l\sqrt{M^2l^2-j^2}\right).
\end{eqnarray}
For the spacetime (\ref{metric2}) after the radial coordinate transformation (\ref{transfor}), we find that the Kretschmann scalar has a form $R^{\mu\nu\rho\sigma}R_{\mu\nu\rho\sigma}=\frac{12}{l^4(1+s)^2}$,
which is the same as in the spacetime (\ref{metric}) without the transformation. This means that the spacetime (\ref{metric2}) has no curvature singularity at the origin as in the spacetime (\ref{metric}). Similarly, the temperatures $T_{\pm}$ of black hole (\ref{metric2})  at the outer ($+$) and inner ($-$) horizons can be expressed as \cite{d2302}
\begin{eqnarray}
T_{\pm}=\pm\frac{r_{\pm}-r_{\mp}}{2\pi l^2\sqrt{1+s}\sqrt{r_{\pm}}}.
\end{eqnarray}
Moreover, we find that the first law of thermodynamics and the Smarr formula for the inner and outer horizons respectively have the forms
\begin{eqnarray}
dE_{\pm} =T_{\pm}dS_{\pm}+\Omega_{\pm}dJ+V_{\pm}dP,\quad\quad\quad 0=T_{\pm}S_{\pm}+\Omega_{\pm}J-2V_{\pm}P,
\end{eqnarray}
with
\begin{eqnarray}
&&E_{\pm}=\frac{1}{8}\bigg(\frac{r_{\pm}}{l^2}+\frac{j^2}{4r_{\pm}}\bigg),\quad\quad\quad S_{\pm}=\int \bigg(\frac{dE_{\pm}}{T_{\pm}}\bigg)_{J,P}=\frac{1}{2}\pi \sqrt{1+s}\sqrt{r_{\pm}},\nonumber\\
&&\Omega_{\pm}=\frac{4J}{r_{\pm}},\quad\quad\quad V_{\pm}=\bigg(\frac{\partial E_{\pm}}{\partial P}\bigg)_{S,J}=\pi (1+s)r_{\pm},\quad\quad\quad P=\frac{1}{8\pi l^2(1+s)},
\end{eqnarray}
where $E_{\pm}$ denote the energy of the black hole at the outer and inner horizons respectively, and are related to the ADM mass $M_{ADM}=\frac{1}{8}M$ by $E_+=M_{ADM}$ and $E_{-}=-M_{ADM}$. The angular momentum is $J=j/8$. These formulas are exactly the same as those for the spacetime (\ref{metric}) without the transformation (\ref{transfor}) \cite{d2302}.
Thus, the radial coordinate transformation (\ref{transfor}) does not change the intrinsic properties of the spacetime (\ref{metric}). However, it will simplify the radial equation of the scalar perturbational field.
In the three dimensional spacetime, the scalar perturbation can be assumed as a form $\psi=e^{-i\omega t+im\theta}R(r)$, where $m$ is a quantum number of the angular coordinate $\theta$, and $\omega$ is the frequency of the scalar field perturbation. Then, the Klein-Gordon equation of the massive scalar field
 \begin{eqnarray}\label{m1}
  \frac{1}{\sqrt{-g}}\partial_{\mu}\bigg(\sqrt{-g}g^{\mu\nu}\partial_{\nu}\psi\bigg)-\mu^2_0\psi=0,
  \end{eqnarray}
in the background of a rotating BTZ-like black hole in the Einstein-bumblebee gravity can be expressed as
  \begin{eqnarray}\label{ra1}
  \frac{d^2R(r)}{dr^2}+\bigg[\frac{1}{r}+\frac{f'(r)}{f(r)}\bigg]\frac{dR(r)}{dr}+\frac{(s+1)}{4}
  \bigg[\frac{1}{rf(r)^2}
  \left(\omega-\frac{mj}{2r}\right)^2-\frac{m^2}{r^2f(r)}-\frac{\mu^2_0}{rf(r)}\bigg]R(r)=0,
  \end{eqnarray}
where the prime is the derivative with respect to the coordinate $r$ and $\mu_0$ is the mass of the scalar perturbational field.
Defining the variable
\begin{eqnarray}
z=\frac{r-r_+}{r-r_-},
\end{eqnarray}
we find that the radial equation (\ref{ra1}) can be rewritten as
 \begin{eqnarray}\label{radial}
 z(1-z)\frac{d^2R(z)}{dz^2}+(1-z)\frac{dR(z)}{dz}+\left(\frac{A}{z}-\frac{B}{1-z}-C\right)R(z)=0,
\end{eqnarray}
with
\begin{eqnarray}
 A&=&\frac{l^4r_+(1+s)(\omega-\frac{jm}{2r_+})^2}{4(r_+-r_-)^2},\quad\quad\quad
 B=\frac{l^2(1+s)\mu^2_0}{4},\quad\quad\quad
 C=\frac{l^4r_-(1+s)(\omega-\frac{jm}{2r_-})^2}{4(r_+-r_-)^2}.
\end{eqnarray}
As in \cite{cjh123,BhattacharjeeSB}, one can use the following redefinition of the radial function
\begin{eqnarray}
 R(z)=z^{\alpha} (1-z)^{\beta} F(z),
\end{eqnarray}
and then the new function $F(z)$ satisfies the Hypergeometric differential equation
 \begin{eqnarray}\label{radial2}
 z(1-z)\frac{d^2F(z)}{dz^2}+[(1+2\alpha)-(1+2\alpha+2\beta)z]\frac{dF(z)}{dz}+\left(\frac{A'}{z}-\frac{B'}{1-z}-C'\right)F(z)=0,
\end{eqnarray}
 with
\begin{eqnarray}
 A'=A+\alpha^2,\quad\quad\quad B'=B+\beta(1-\beta),\quad\quad\quad C'=C+(\alpha+\beta)^2.
\end{eqnarray}
To find the solution for the total physical spacetime, one must remove the poles at $z=0$ and $z=1$ by applying some constraints on the coefficient of function $F$, i.e.,
\begin{eqnarray}
 A'=0,\quad\quad\quad B'=0.
\end{eqnarray}
These mean that
\begin{eqnarray}
\alpha&=&\mp i\sqrt{A}=\mp i\frac{l^2\sqrt{r_+}\sqrt{1+s}(\omega-\frac{jm}{2r_+})}{2(r_+-r_-)}, \quad\quad\quad
 \beta=\frac{1}{2}\left(1\pm\sqrt{1+4B}\right).
\end{eqnarray}
It implies that the corresponding Breitenlohner-Freedman bound $\mu^2_{0BF}$ \cite{BFbound} in the spacetime (\ref{metric2}) becomes as
\begin{eqnarray}
\mu^2_{0BF}l^2=-\frac{1}{(1+s)}.
\end{eqnarray}
Only the scalar field with the mass $\mu_0$ above the Breitenlohner-Freedman bound does not induce an instability in the spacetime. It is clear that with the increase of the spontaneous Lorentz symmetry breaking parameter $s$, the Breitenlohner-Freedman bound increases and the corresponding range of the mass $\mu_0$ of scalar field yielding instability becomes narrow.
The equation (\ref{radial2}) can finally reduce to the Hypergeometric equation
\begin{eqnarray}\label{radial3}
 z(1-z)\frac{d^2F(z)}{dz^2}+[c-(1+a+b)z]\frac{dF(z)}{dz}+ab F(z)=0,
\end{eqnarray}
with
\begin{eqnarray}
 c&=&2\alpha+1, \quad\quad\quad
 a=\alpha+\beta+i\sqrt{C},\quad\quad\quad
 b=\alpha+\beta-i\sqrt{C}.\label{b1}
\end{eqnarray}
Thus, the general solution of equation (\ref{radial}) can be given by a linear combination of Hypergeometric functions $F$, i.e.,
\begin{eqnarray}\label{hype1}
 R=z^{\alpha} (1-z)^{\beta}[C_1F(a,b,c;z)+C_2 z^{1-c}F(a-c+1,b-c+1,2-c;z)],
\end{eqnarray}
where $C_1$ and $C_2$ are the constants of integration.

Given that the wave functions are purely ingoing at the horizon $z=0$, the
boundary conditions determine $\alpha=- i\sqrt{A}$. Thus, in the near horizon region, the
contribution to the wave function only comes from the first term in Eq. (\ref{hype1}), i.e.,
\begin{eqnarray}\label{nearhorizon}
 R=C_1z^{\alpha} (1-z)^{\beta}F(a,b,c;z).
\end{eqnarray}
In the far region $r\gg r_+$, one can make use of the property of the hypergeometric function \cite{mb}
\begin{eqnarray}
F(a,b,c;z)&=&\frac{\Gamma(c)\Gamma(c-a-b)}{\Gamma(c-a)\Gamma(c-b)}
F(a, b, a+b-c+1; 1-z)\nonumber\\
&+&(1-z)^{c-a-b}\frac{\Gamma(c)\Gamma(a+b-c)}{\Gamma(a)\Gamma(b)}
F(c-a, c-b, c-a-b+1; 1-z),\label{r2}
\end{eqnarray}
and obtain the asymptotic behavior of the wave function
$R$ at the spatial infinity (i.e., $z\rightarrow1$)
 \begin{eqnarray}
 R\simeq z^{\alpha} (1-z)^{\beta}
 \frac{\Gamma(c)\Gamma(c-a-b)}{\Gamma(c-a)\Gamma(c-b)}.
 \end{eqnarray}
In the rotating BTZ-like black hole for the Einstein-bumblebee gravity (\ref{metric}), the effective potential in the radial equation for the scalar perturbation tends to infinity as $r\rightarrow\infty$, so the physical
requirement should be imposed as in Refs. \cite{wbq1,wbq2}, i.e., the wavefunction is
just purely outgoing at spatial infinity and its corresponding flux is finite. After some careful analysis, one can find that all of the divergent terms in the flux are proportional to
\begin{eqnarray}
   \left |\frac{\Gamma(c)\Gamma(c-a-b)}{\Gamma(c-a)\Gamma(c-b)}\right
   |^2.
\end{eqnarray}
Thus, the boundary condition of the non-divergent flux at the asymptotic infinity leads to
\begin{eqnarray}
  c-a=-n, \hspace{5ex} \mbox{or}\hspace{5ex} c-b=-n,
\end{eqnarray}
with $n$ being a non-negative integer. It should be noted that these two relations can be also obtained by simply imposing vanishing Dirichlet condition at spatial infinity.

From the relation $c-a=-n$, we find that the right-moving quasinormal frequency obeys to
 \begin{eqnarray}\label{case1}
 -\frac{il^2\sqrt{1+s}}{2(\sqrt{r_+}+\sqrt{r_-})}\omega-\frac{ijl^2 m\sqrt{1+s}}{4\sqrt{r_+r_-}(\sqrt{r_+}+\sqrt{r_-})}
 +\frac{1}{2}\left(1+\sqrt{1+4B}\;\right)=-n.
 \end{eqnarray}
Solving Eq. (\ref{case1}), one can obtain the right-moving quasinormal frequency  for the scalar field in the background of the  rotating BTZ-like black hole for the Einstein-bumblebee gravity
\begin{eqnarray}\label{qraR}
 \omega_R&=&-\frac{m}{l}-i\frac{2(\sqrt{r_+}+\sqrt{r_-})}{l^2\sqrt{1+s}}\bigg[
 n+\frac{1}{2}+\frac{1}{2}\sqrt{1+l^2(1+s)\mu^2_0}\bigg].
 \end{eqnarray}
Similarly, from the condition $c-b=-n$, we obtain the left-moving quasinormal frequency
\begin{eqnarray}\label{qraL}
 \omega_L&=&\frac{m}{l}-i\frac{2(\sqrt{r_+}-\sqrt{r_-})}{l^2\sqrt{1+s}}\bigg[
 n+\frac{1}{2}+\frac{1}{2}\sqrt{1+l^2(1+s)\mu^2_0}\bigg].
 \end{eqnarray}
\begin{figure}
\includegraphics[width=7.6cm]{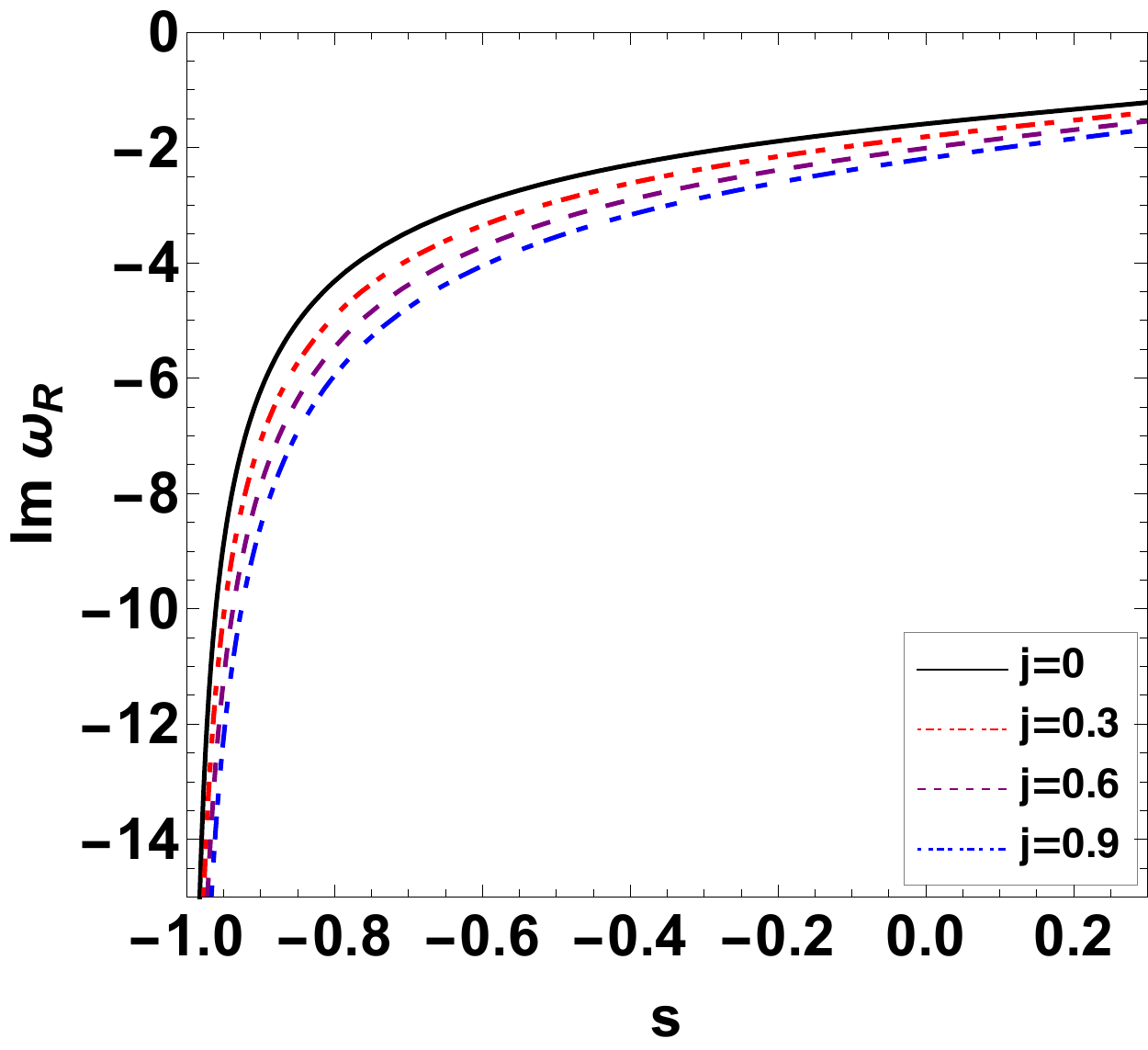}\quad\ \quad\includegraphics[width=7.6cm]{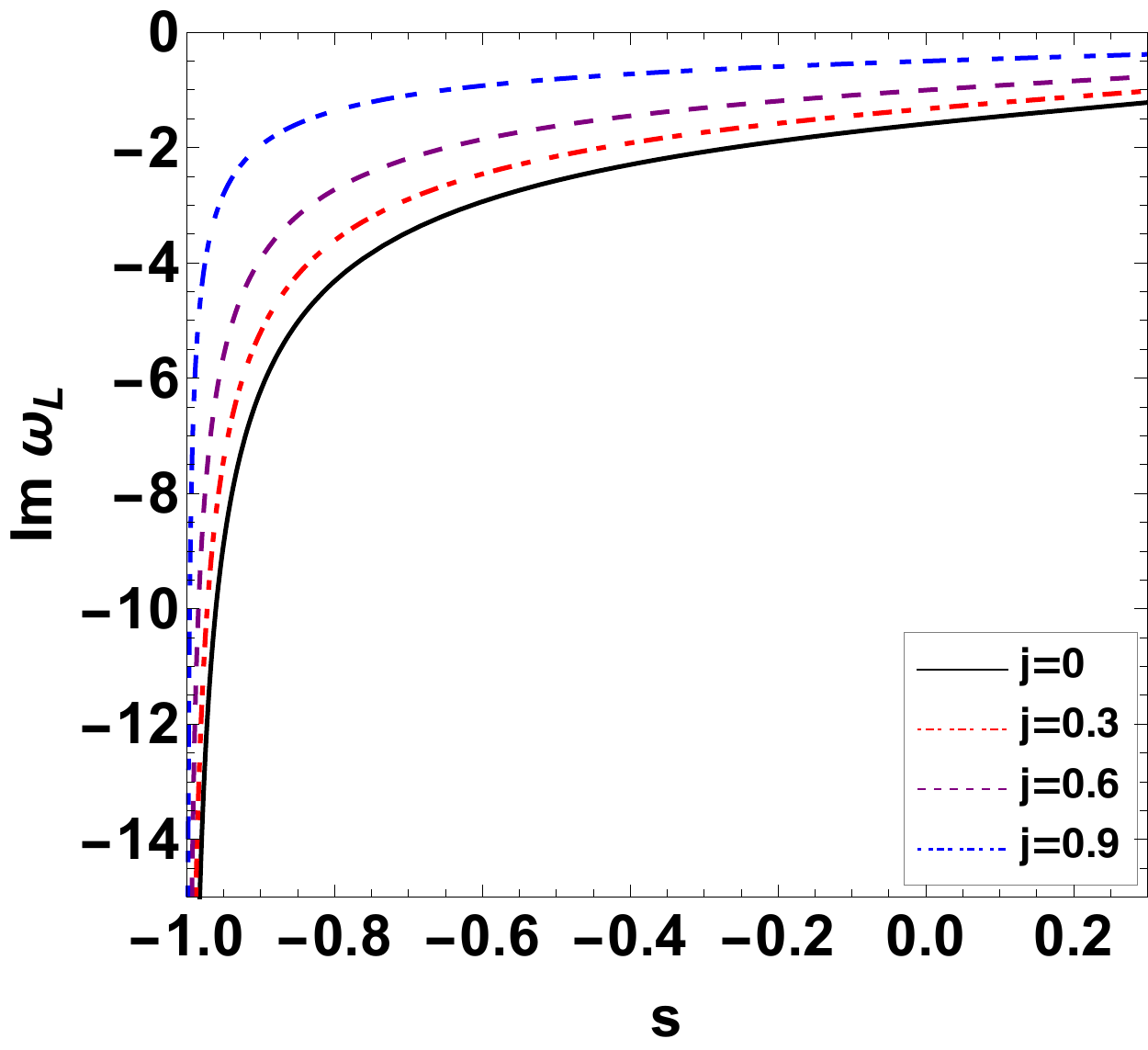}
\includegraphics[width=7.6cm]{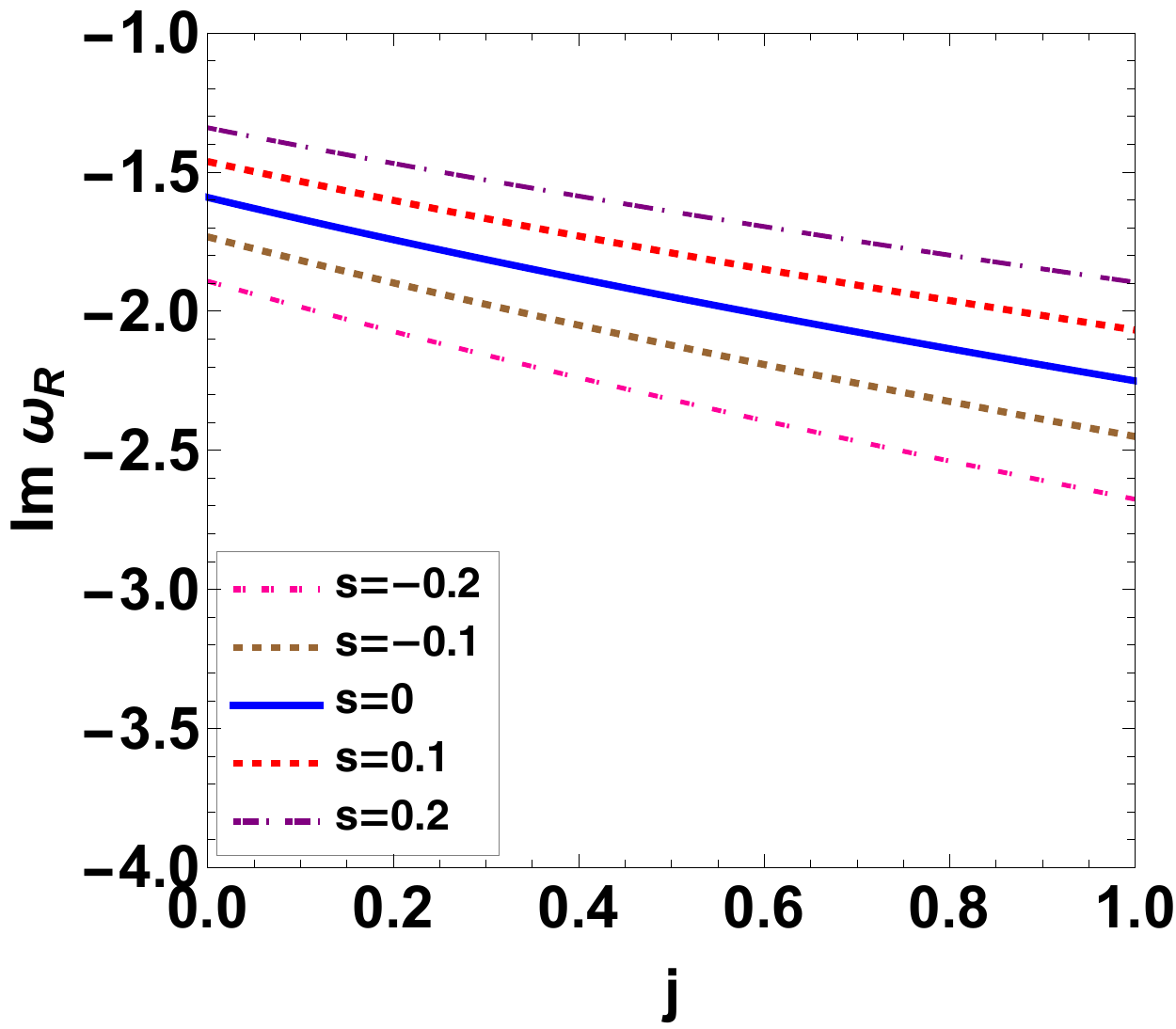}\quad\ \quad\includegraphics[width=7.9cm]{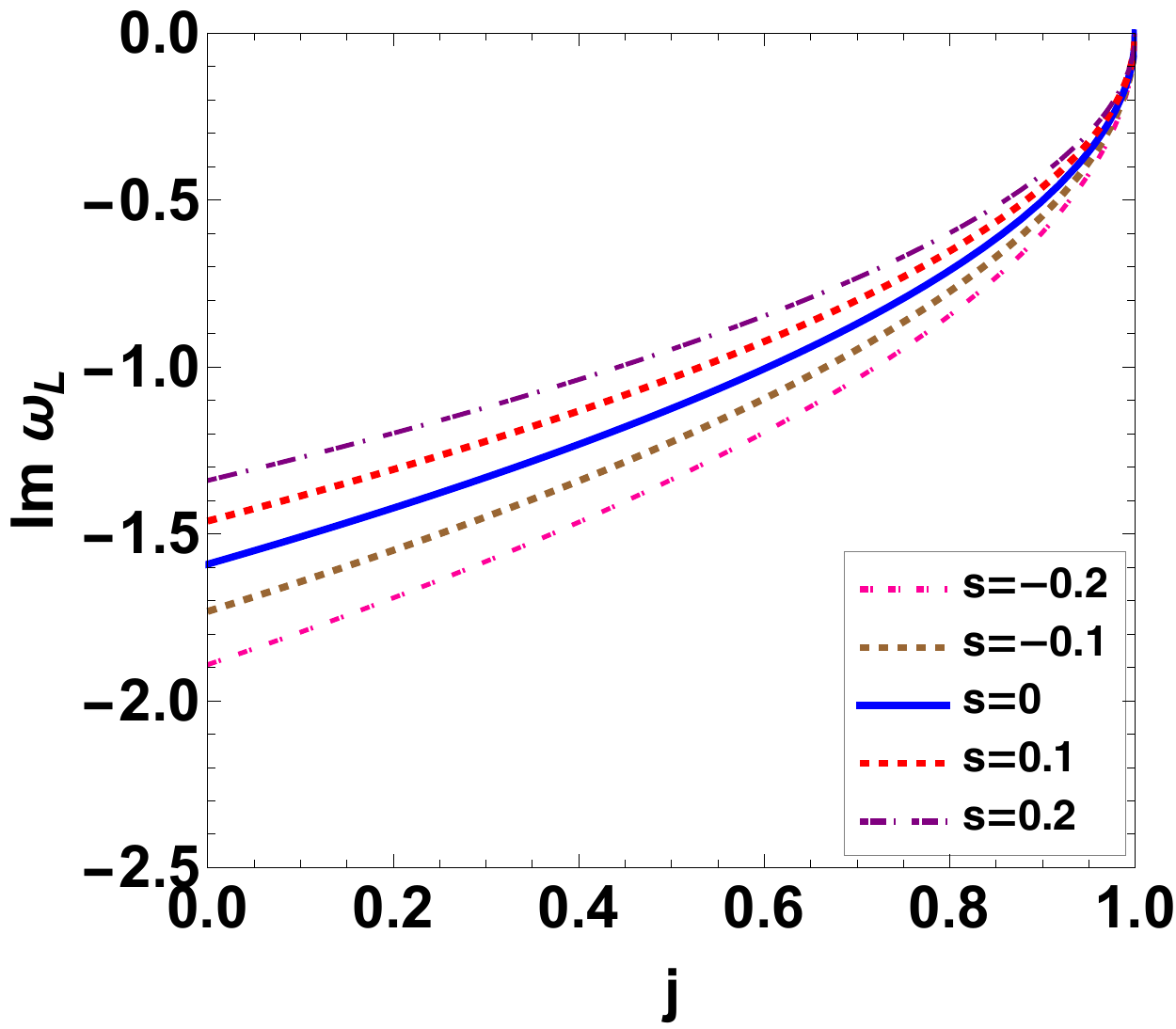}
\caption{Variety of the imaginary parts of right-moving and left-moving quasinormal
frequencies with the  symmetry breaking parameter $s$ and the spin parameter $j$. Here, we set $M=1$, $l=1$ and $\mu^2_0=-0.65$. }
\label{wsj}
\end{figure}
Obviously, the real parts of both right-moving and left-moving quasinormal
frequencies are defined by only the scalar field's angular quantum number $m$ and the cosmological  parameter $l$. However, the imaginary parts depend on the spontaneous Lorentz symmetry breaking parameter $s$ and the spin parameter $j$. Comparing with the case without the Lorentz symmetry breaking, the positive symmetry breaking parameter $s$ makes the absolute values of the imaginary parts decrease, but the negative $s$ makes them increase. Thus, the scalar perturbation field decays more rapidly in the case with the negative breaking parameter. Moreover, with the increase of the spin parameter $j$, we find that the absolute values of the imaginary parts increase for the right-moving modes, but decrease for the left-moving modes.  These changes of the imaginary parts of quasinormal frequencies are also shown in Fig. \ref{wsj}.
As the Lorentz symmetry breaking parameter $s$  vanishes, one can find that the formulas (\ref{qraR})
and (\ref{qraL}) of quasinormal frequencies are consistent with those of
the scalar field in the usual rotating BTZ black hole spacetime.

Finally, we will probe the relationship between the $AdS/CFT$ correspondence and the quasinormal modes of the scalar field in the background of the rotating BTZ-like black hole (\ref{metric2}) in the Einstein-bumblebee gravity. At thermal equilibrium of the black hole, the two sectors may have different temperatures $(T_L, T_R)$ for the rotating  BTZ-like black holes (\ref{metric2}), i.e., \cite{d2302}
\begin{eqnarray}\label{Temper}
 T_L=\frac{\sqrt{r_+}-\sqrt{r_-}}{2\pi l^2\sqrt{1+s}},\quad\quad\quad\quad T_R=\frac{\sqrt{r_+}+\sqrt{r_-}}{2\pi l^2\sqrt{1+s}}.
 \end{eqnarray}
From Eqs. (\ref{qraR}) and (\ref{qraL}), we can find that the quasinormal frequencies of
a scalar field around the BTZ-like  black hole in the Einstein-bumblebee gravity
can be rewritten as
\begin{eqnarray}\label{wLR}
 \omega_L&=&\frac{m}{l}-4\pi iT_L(n+h_L),\quad\quad\quad\quad\omega_R=-\frac{m}{l}-4\pi iT_R(n+h_R),
 \end{eqnarray}
with the conformal weights of its corresponding operator in the dual CFT
\begin{eqnarray}
h_L=h_R=\frac{1}{2}+\frac{1}{2}\sqrt{1+l^2(1+s)\mu^2_0}.
 \end{eqnarray}
Thus, one can find that the conjectured $AdS/CFT$ correspondence still holds for the BTZ-like black hole in the Einstein-bumblebee gravity.  Moreover, we also find that the conformal weights $h_L$ and $h_R$ increase with the Lorentz symmetry breaking parameter $s$, but are independent of the spin parameter of the black hole.

From Eq. (\ref{wLR}), one can find that the quasinormal frequencies are related to thermodynamic properties of the black hole. The literature \cite{masslessbh} shows that the left-moving and right-moving modes for the scalar field become normal modes without any decay in a static and massless AdS spacetime and then the quasinormal frequencies can reflect the transition between black hole phase and massless phase.  Repeating the similar operations, it is easy to obtain that there only exists the normal mode with $ \omega_{L,R}=\pm \frac{m}{l}$ in the spacetime (\ref{metric2}) with $M=0$ and $j=0$. For the rotating Einstein-bumblebee BTZ-like hole, one can find that the mass parameter $M$ must satisfy $M\geq j/l$ and in the extremal black hole case the mass parameter $M$ is equal to its minimum  $M=j/l$. For the extremal rotating  Einstein-bumblebee BTZ-like hole  the radial equation (\ref{ra1}) of the scalar perturbation can be expressed as
\begin{eqnarray}
&&(2r-Ml^2)^4\frac{d^2R(r)}{dr^2}+4(2r-Ml^2)^3\frac{dR(r)}{dr}\nonumber\\&&+(s+1)l^2
  \bigg\{4(\omega l-m)[(\omega l+m)r-ml^2M]-(2r-Ml^2)^2\mu^2_0\bigg\}R(r)=0.
  \label{eqphase1}
\end{eqnarray}
Redefining
\begin{eqnarray}
z=\frac{Ml^2}{2r-Ml^2},
\end{eqnarray}
Eq. (\ref{eqphase1}) becomes
\begin{eqnarray}
\frac{d^2R(z)}{dz^2}+\frac{(s+1)l}{4}
  \bigg[\frac{2(\omega l-m)^2}{M}+\frac{(\omega^2 l^2-m^2)}{Mz}-\frac{\mu^2_0l^2}{z^2}\bigg]R(z)=0.
  \label{eqphase2}
\end{eqnarray}
The general solution of the above equation is given by
\begin{eqnarray}\label{hype1New}
 R=\mathcal{A}_1(2z)^{\beta_1}e^{-\zeta_{-} z} F[\beta_1-\zeta_+,\;2\beta_1;\;2\zeta_-z]+\mathcal{A}_2(2z)^{1-\beta_1}e^{-\zeta_{-} z} F[1-\beta_1-\zeta_+,\;2(1-\beta_1);\;2\zeta_-z],
\end{eqnarray}
with
\begin{eqnarray}
\zeta_{\pm}=\frac{2i \sqrt{1+s}(m\pm\omega l)}{\sqrt{2M}},\quad\quad\quad \beta_1=\frac{1}{2}\bigg[1+\sqrt{1+l^2(1+s)\mu^2_0}\bigg],
\end{eqnarray}
where $F[a,b;z]$ is the confluent hypergeometric function. $\mathcal{A}_1$ and $\mathcal{A}_2$ are the constants of integration.
As in \cite{extremalbh}, the boundary condition of no outgoing wave at the spatial infinity requires $\mathcal{A}_2$ in Eq. (\ref{hype1New}) must be zero and the asymptotic limit of the wave function becomes
\begin{eqnarray}
\psi_{\infty}=\mathcal{A}_1e^{-i\omega t+im\theta}e^{-\frac{\zeta_{-}}{2r}} \bigg(\frac{2}{r}\bigg)^{\beta_1}.
\end{eqnarray}
Thus, the conserved radial current of the scalar perturbation at the spatial infinity becomes
\begin{eqnarray}
\mathcal{J}=\psi^{*}\frac{d\psi}{dr}-\psi\frac{d\psi^{*}}{dr}=0,
\end{eqnarray}
and the corresponding flux $\mathcal{F}=\sqrt{-g}\mathcal{J}/(2i)$ also vanishes at the spatial infinity. This means the absence of quasinormal modes for the extremal rotating  Einstein-bumblebee BTZ black hole under the scalar perturbation. Thus, there also only exists the normal mode without imaginary part in the spacetime (\ref{metric2}) with $M=j/l$. This property is valid for the arbitrary Lorentz symmetry breaking parameter $s$. However, in the non-extremal Einstein-bumblebee BTZ black hole case, the quasinormal frequencies (\ref{qraR}) and (\ref{qraL}) have the non-zero imaginary parts for the left-moving and right-moving modes. Therefore, combining with the previous discussion for the case $j=0$, it is easy to obtain that the change of the imaginary parts of the left-moving and right-moving quasinormal frequencies from a non-zero value to zero reflects a kind of phase transitions from the rotating non-extremal BTZ-like black hole to the extremal black hole or from the static BTZ-like black hole to the massless AdS phase.

Actually, as Cai \textit{et al} observed in \cite{cairg}, for the BTZ-like spacetime (\ref{metric2}),  some second moment related to the Hawking temperature $T_H=1/\beta$ (for example, $\langle\delta\beta\delta\beta\rangle \sim \sqrt{1+s}(M l-j)^{-3/2}$) in the microcanonical ensemble diverges as $M\rightarrow j/l$. The similar behaviors also appear in the previous nonrotating case as $M\rightarrow0$. Therefore, the above phase transition belongs to the thermodynamic phase transition. As in \cite{cairg,Kaburaki}, for such kind of phase transitions, the difference between $r_+$ and $r_-$ can be served as an order parameter $\epsilon\equiv r_+-r_-$. From Eq. (\ref{qraL}), it is easy to find that the imaginary part of quasinormal frequencies for the left-moving mode is directly related to the order parameter $\epsilon$ by
\begin{eqnarray}\label{imwLR}
 {\rm Im}(\omega_L)=-\frac{2(n+h_L)\epsilon}{l^2\sqrt{1+s}(\sqrt{r_+}+\sqrt{r_-})}.
 \end{eqnarray}
Although the quasinormal frequencies for the right-moving mode have not such direct connection with the order parameter $\epsilon$, their values are non-zero in the non-extremal BTZ-like black hole phase and are zero in the extremal black hole phase. In this sense, the change of the imaginary parts of the right-moving quasinormal frequencies also reflects such kind of thermodynamic phase transitions between BTZ-like black holes.

\section{Summary}

We have studied the quasinormal modes of a scalar perturbation around a rotating BTZ-like black hole in the Einstein-bumblebee gravity. We find that the spontaneous Lorentz symmetry breaking parameter imprints only in the imaginary parts of the quasinormal frequencies for the right-moving and left-moving modes. Comparing with the case without the Lorentz symmetry breaking, the positive symmetry breaking parameter $s$ makes the absolute values of the imaginary parts decrease, but the negative $s$ makes them increase. Thus, the scalar perturbation field decays more rapidly in the case with the negative Lorentz symmetry breaking parameter.  Moreover, with the increase of the spin parameter $j$, we find that the absolute values of the imaginary parts increase for the right-moving modes, but decrease for the left-moving modes. The real parts depend on only the scalar field's angular quantum number $m$ and the cosmological  parameter $l$, which is consistent with the quasinormal modes in the usual BTZ black hole.

Moreover, we discuss the $AdS/CFT$ correspondence from the quasinormal modes and find that it still holds for the BTZ-like black hole in the Einstein-bumblebee gravity. The Lorentz symmetry breaking parameter enhances the left and right conformal weights $h_L$ and $h_R$ of the operators dual to the scalar field in the boundary. In addition, the quasinormal frequencies are related to thermodynamic properties of the black hole and the change of the imaginary parts of the quasinormal frequencies reflects a kind of phase transitions from the nonrotating BTZ-like black hole to the massless AdS phase or from the rotating BTZ-like black hole to the extremal AdS black hole. These results could be helpful to understand the $AdS/CFT$ correspondence and the Einstein-bumblebee gravity with the Lorentz symmetry violation.

\begin{acknowledgments}

This work was supported by the National Key Research and Development Program of China (Grant No. 2020YFC2201400) and National Natural Science Foundation of China (Grant Nos. 12275079 and 12035005).

\end{acknowledgments}


\begin{thebibliography}{99}
\baselineskip=0.5cm

\bibitem{lv01} G.T. Zatsepin and V.A. Kuzmin, \textit{Upper limit of the spectrum of cosmic rays}, JETP Lett. {\bf4},
78 (1966).

\bibitem{lv02} M. Takeda \emph{et al.}, \textit{Extension of the cosmic ray energy spectrum beyond the predicted
Greisen-Zatsepin-Kuz'min cutoff}, Phys. Rev. Lett. {\bf81}, 1163 (1998); arXiv:astro-ph/9807193.

\bibitem{lv02gc} V.A. Kostelecky and S. Samuel, \textit{Photon and graviton masses in string theories}, Phys. Rev. Lett. {\bf66}, 1811 (1991).

\bibitem{lv02gc1} V.A. Kostelecky and R. Potting, \textit{CPT, strings, and meson factories}, Phys. Rev. D {\bf51}, 3923 (1995).

\bibitem{lv02gc2} D. Colladay and V.A. Kostelecky, \textit{CPT violation and the standard model}, Phys. Rev. D {\bf55}, 6760 (1997).

\bibitem{string1} V.A. Kostelecky and S. Samuel, \textit{Spontaneous breaking of Lorentz symmetry in string theory}, Phys. Rev. D {\bf39}, 683 (1989).

\bibitem{lvbh2} V.A. Kostelecky and S. Samuel, \textit{Gravitational Phenomenology in Higher Dimensional Theories and Strings}, Phys. Rev. D {\bf40}, 1886 (1989).

\bibitem{lvbh2s1} R. Bluhm and V.A. Kostelecky, \textit{Spontaneous Lorentz violation, Nambu-Goldstone modes, and gravity}, Phys. Rev. D {\bf71}, 065008 (2005); arXiv:hep-th/0412320.


\bibitem{lvbhh1} O. Bertolami and J. Paramos, \textit{The Flight of the bumblebee: Vacuum solutions of a gravity
model with vector-induced spontaneous Lorentz symmetry breaking}, Phys. Rev. D {\bf72}, 044001 (2005); arXiv:hep-th/0504215.

\bibitem{lvbhh2} Q.G. Bailey and V.A. Kostelecky, \textit{Signals for Lorentz violation in post-Newtonian gravity},
Phys. Rev. D {\bf74}, 045001 (2006); arXiv:gr-qc/0603030.

\bibitem{lvbhh12} R. Bluhm, N.L. Gagne, R. Potting, and A. Vrublevskis, \textit{Constraints and Stability in Vector Theories with Spontaneous Lorentz Violation}, Phys. Rev. D {\bf 77}, 125007 (2008); Erratum ibid. {\bf 79} 029902 (2009); arXiv:0802.4071 [hep-th].

\bibitem{lvbhh3} V.A. Kostelecky and J. Tasson, \textit{Prospects for Large Relativity Violations in Matter-Gravity
Couplings}, Phys. Rev. Lett. {\bf 102}, 010402 (2009); arXiv:0810.1459 [gr-qc].

\bibitem{lvbhh4} M.D. Seifert, \textit{Generalized bumblebee models and Lorentz-violating electrodynamics},
Phys. Rev. D {\bf81}, 065010 (2010); arXiv:0909.3118 [hep-ph].

\bibitem{lvbhh5} R.V. Maluf, C.A.S. Almeida, R. Casana, and M. Ferreira, \textit{Einstein-Hilbert graviton modes
modified by the Lorentz-violating bumblebee Field}, Phys. Rev. D {\bf90}, 025007 (2014); arXiv:1402.3554 [hep-th].

\bibitem{lvbhh6} J. P\'{a}ramos and G. Guiomar, \textit{Astrophysical Constraints on the Bumblebee Model},
Phys. Rev. D {\bf90}, 082002 (2014); arXiv:1409.2022 [astro-ph].

\bibitem{lvbhh7} C.A. Escobar and A. Mart\'{i}n-Ruiz, \textit{Equivalence between bumblebee models and
electrodynamics in a nonlinear gauge}, Phys. Rev. D {\bf95}, 095006 (2017); arXiv:1703.01171 [hep-th].

\bibitem{lvbhh8} J.F. Assun\~{a}o, T. Mariz, J.R. Nascimento, and A.Y. Petrov, \textit{Dynamical Lorentz symmetry
breaking in a tensor bumblebee model}, Phys. Rev. D {\bf100}, 085009 (2019); arXiv:1902.10592 [hep-th].

\bibitem{lvbh10} D. Capelo and J. P\'{a}ramos, \textit{Cosmological implications of Bumblebee vector models}, Phys. Rev. D {\bf91}, 104007 (2015); arXiv:1501.07685 [gr-qc].

\bibitem{ReyesSS} C.M. Reyes, M. Schreck, and A. Soto, \textit{Cosmology in the presence of diffeomorphism-violating, nondynamical background fields}, Phys. Rev. D  {\bf 106}, 023524 (2022); arXiv:2205.06329 [gr-qc].

\bibitem{UniyalKS} A. Uniyal, S. Kanzi, and \.{I}. Sakall{\i}, \textit{Greybody factors of bosons and fermions emitted from higher dimensional dS/AdS black holes in Einstein-bumblebee gravity theory}, arXiv:2207.10122 [hep-th].

\bibitem{KhodadiS} M. Khodadi and M. Schreck, \textit{Hubble tension as a guide for refining the early Universe: Cosmologies with explicit local Lorentz and diffeomorphism violation}, Phys. Dark Universe  {\bf 39}, 101170 (2023).

\bibitem{Fang} W. Liu, X. Fang, J. Jing, and J. Wang, \textit{Exact Kerr-like solution and its shadow in a gravity model with spontaneous Lorentz symmetry breaking}, Eur. Phys. J. C  {\bf 83}, 83 (2023); arXiv:2211.03156 [gr-qc].


\bibitem{lvbh1}R. Casana, A. Cavalcante, F.P. Poulis, and E.B. Santos, \textit{Exact Schwarzschild-like solution in a bumblebee gravity model}, Phys. Rev. D {\bf97}, 104001 (2018); arXiv:1711.02273 [gr-qc].

\bibitem{lvbh3}A. Ovgun, K. Jusufi, and I. Sakalli, \textit{Gravitational Lensing Under the Effect of Weyl and
Bumblebee Gravities: Applications of Gauss-Bonnet Theorem}, Ann. Phys. {\bf399}, 193 (2018); arXiv:1805.09431 [gr-qc].

\bibitem{lvbh5}R. Oliveira, D.M. Dantas, and C.A.S. Almeida, \textit{Quasinormal frequencies for a black hole in a bumblebee gravity},  EPL {\bf135}, 10003 (2021); arXiv:2105.07956 [gr-qc].

\bibitem{lvbh4}S. Kanzi and I. Sakalli, \textit{GUP Modified Hawking Radiation in Bumblebee Gravity},
Nucl. Phys. B {\bf946}, 114703 (2019); arXiv:1905.00477 [hep-th].

\bibitem{lvbh6} I. G\"{u}ll\"{u} and A. \"{O}vg\"{u}n, \textit{Schwarzschild-like black hole with a topological defect in bumblebee gravity},  Ann. Phys. {\bf436}, 168721 (2022); arXiv:2012.02611 [gr-qc].

\bibitem{lvbh7} R.V. Maluf and J.C.S. Neves, \textit{Black holes with a cosmological constant in bumblebee gravity}, Phys. Rev. D {\bf103}, 044002 (2021).

\bibitem{lvbh8} C. Ding, X. Chen, and X. Fu, \textit{Einstein-Gauss-Bonnet gravity coupled to bumblebee field in four dimensional spacetime}, Nucl. Phys. B {\bf 975},  115688 (2022); arXiv:2102.13335 [gr-qc].

\bibitem{lvbh9} A. \"{O}vg\"{u}n, K. Jusufi, and I. Sakall, \textit{Exact traversable wormhole solution in bumblebee gravity}, Phys. Rev. D {\bf99}, 024042 (2019); arXiv:1804.09911 [gr-qc].

\bibitem{lvbhrot1} C. Ding, C. Liu, R. Casana, and A. Cavalcante, \textit{Exact Kerr-like solution and its shadow in a gravity model with spontaneous Lorentz symmetry breaking}, Eur. Phys. J. C {\bf 80}, 178 (2020); arXiv:1910.02674 [gr-qc].

\bibitem{lvbhrot1s}H. Wang and S. Wei, \textit{Shadow cast by Kerr-like black hole in the presence of plasma in Einstein-bumblebee gravity},  Eur. Phys. J. Plus {\bf 137}, 571 (2022); arXiv:2106.14602 [gr-qc].


\bibitem{lvbhrot2} C. Liu, C. Ding, and J. Jing, \textit{Thin accretion disk around a rotating Kerr-like black hole in Einstein-bumblebee gravity model}, arXiv:1910.13259 [gr-qc].

\bibitem{lvbhrot3} R. Jiang, R. Lin, and X. Zhai, \textit{Superradiant instability of the Kerr-like black hole in Einstein-bumblebee gravity}, Phys. Rev. D {\bf 104}, 124004  (2021); arXiv:2108.04702 [gr-qc].

\bibitem{lvbhrot4} Z. Li and A. \"{O}vg\"{u}n, \textit{Finite-distance gravitational deflection of massive particles by a
Kerr-like black hole in the bumblebee gravity model}, Phys. Rev. D {\bf 101},  024040 (2020); arXiv:2001.02074 [gr-qc].

\bibitem{lvbhrot5} S.K. Jha and A. Rahaman, \textit{Bumblebee gravity with a Kerr-Sen-like solution and its Shadow}, Eur. Phys. J. C {\bf81}, 345 (2021); arXiv:2011.14916 [gr-qc].

\bibitem{RPM1} S.E. Motta, T.M. Belloni, L. Stella, T. Muoz-Darias, and R. Fender, \textit{Precise mass and spin measurements for a stellar-mass black hole through X-ray timing: the case of GRO J1655-40}, Mon. Not. Roy. Astron. Soc. {\bf 437}, 2554 (2014); arXiv:1309.3652 [astro-ph].

\bibitem{XTE1} J.A. Orosz, J.F. Steiner, J.E. McClintock, M.A.P. Torres, R.A. Remillard, C.D. Bailyn, and J.M. Miller, \textit{An Improved Dynamical Model for the Microquasar XTE J1550-564}, Astrophys. J. {\bf 730}, 75 (2011); arXiv:1101.2499 [astro-ph].

\bibitem{GRS1} M.J. Reid, J.E. McClintock, J.F. Steiner, D. Steeghs, R.A. Remillard, V. Dhawan, and R. Narayan, \textit{A Parallax Distance to the Microquasar GRS 1915+105 and a Revised Estimate of its Black Hole Mass}, Astrophys. J. {\bf 796}, 2 (2014); arXiv:1409.2453 [astro-ph].

\bibitem{test1}  Z. Wang, S. Chen, and J. Jing, \textit{Constraint on parameters of a rotating black hole in Einstein-bumblebee theory by quasi-periodic oscillations}, Eur. Phys. J. C  {\bf 82}, 528 (2022).

\bibitem{d2302} C. Ding, Y. Shi, J. Chen, Y. Zhou, C. Liu, and Y. Xiao, \textit{Rotating BTZ-like black hole and central charges in Einstein-bumblebee gravity}, Eur. Phys. J. C  {\bf 83}, 573 (2023); arXiv:2302.01580 [gr-qc].

\bibitem{BFbound} P. Breitenlohner and D. Z. Freedman, \textit{Stability In Gauged Extended Supergravity}, Annals Phys. {\bf144}, 249 (1982).

\bibitem{AdSCFT1} J. Maldacena, \textit{The Large N limit of superconformal field theories and supergravity}, Adv. Theor. Math. Phys. {\bf 2}, 231 (1998).

\bibitem{AdSCFT2} S.S. Gubser, I.R. Klebanov, and A. M. Polyakov, \textit{Gauge theory correlators from noncritical string theory}, Phys. Lett. B {\bf 428}, 105 (1998).

\bibitem{AdSCFT3} E. Witten, \textit{Anti-de Sitter space and holography}, Adv. Theor. Math. Phys. {\bf 2}, 253 (1998).

\bibitem{cjh123} D. Yekta and M. Shariat, \textit{Propagation of a scalar field with nonminimal
coupling in three dimensions: Hawking radiation and quasinormal modes}, Class. Quantum Grav. {\bf36}, 185005 (2019).

\bibitem{BhattacharjeeSB} S. Bhattacharjee, S. Sarkar, and A. Bhattacharyya, \textit{Scalar perturbations of black holes in Jackiw-Teitelboim gravity}, Phys. Rev. D {\bf 103},  024008 (2021); arXiv:2011.08179 [gr-qc].

 \bibitem{mb} M. Abramowitz and I. Stegun, \textit{Handbook of Mathematical Functions} (Academic, New York, 1996).

\bibitem{wbq1}  B. Chen and Z.B. Xu, \textit{Quasi-normal modes of warped black holes and warped AdS/CFT correspondence}, J. High Energy Phys. {\bf 0911}, 091 (2009).

\bibitem{wbq2}  B. Chen and Z.B. Xu, \textit{Quasinormal modes of warped $AdS_{3}$ black holes and AdS/CFT correspondence}, Phys. Lett. B {\bf 675}, 246 (2009); arXiv:0901.3588 [hep-th].

\bibitem{masslessbh} X. Rao, B. Wang, and G. Yang, \textit{Quasinormal modes and phase Transition of black holes}, Phys. Lett. B {\bf649}, 472 (2007).
\bibitem{extremalbh} J. Cris\'{o}stomo, S. Lepe, and J. Saavedra, \textit{Quasinormal modes of extremal BTZ black hole}, Class. Quant. Grav. {\bf21},  2801 (2004).
\bibitem{cairg} R. Cai, Z. Lu, and Y. Zhang, \textit{Critical behavior in $2+1$ dimensional black holes}, Phys. Rev. D {\bf55}, 853 (1997).
\bibitem{Kaburaki} O. Kaburaki, \textit{Critical behavior of extremal Kerr-Newman black holes}, Gen. Rel. Grav. {\bf28}, 843 (1996).
\end{thebibliography}
\end{document}